\documentclass{article}
\usepackage{spconf,amsmath,graphicx,hyperref}

\usepackage{cite}
\usepackage{xurl}
\usepackage{hyperref}
\usepackage{graphicx}
\usepackage{textcomp}
\usepackage{xcolor}
\usepackage{url}
\usepackage{booktabs}
\usepackage{amsfonts}
\usepackage{nicefrac}
\usepackage{microtype}
\usepackage{subcaption}
\usepackage{array}
\usepackage{algorithm}
\usepackage{algorithmicx}
\usepackage{algpseudocode}
\usepackage{amsmath}
\usepackage{amssymb}
\usepackage{mathtools}
\usepackage{amsthm}
\usepackage{extpfeil}
\usepackage{multirow}
\usepackage{tikz}
\usepackage{makecell}
\usepackage{siunitx}

\usetikzlibrary{quantikz2}
\usetikzlibrary{calc, decorations.pathreplacing}
\usetikzlibrary{positioning, fit, backgrounds, arrows.meta}

\usepackage{thmtools}
\usepackage{mathrsfs}
\usepackage{tabularx}
\usepackage[most]{tcolorbox}
\usepackage{enumerate}
\usepackage{paralist}
\usepackage{enumitem}

\DeclarePairedDelimiterX{\norm}[1]{\lVert}{\rVert}{#1}

\theoremstyle{definition}

\theoremstyle{remark}

\title{Multivariate Time Series Forecasting with Adaptive Non-Local Observables}
\name{
  Yu-Ting Lee$^{1}$ \quad
  Huan-Hsin Tseng$^{2}$ \quad
  Samuel Yen-Chi Chen$^{3}$\thanks{The views expressed in this article are those of the authors and do not represent the views of Wells Fargo. This article is for informational purposes only. Nothing contained in this article should be construed as investment advice. Wells Fargo makes no express or implied warranties and expressly disclaims all legal, tax, and accounting implications related to this article.}
}
\address{
  $^{1}$Graduate Institute of Communication Engineering, National Taiwan University, Taipei, Taiwan \\
  $^{2}$Brookhaven National Laboratory, AI \& ML Department, Upton, NY, USA \\
  $^{3}$Wells Fargo, New York, NY, USA \\
  r14942088@ntu.edu.tw, htseng@bnl.gov, yen-chi.chen@wellsfargo.com
  }

\begin{document}
\maketitle
\begin{abstract}
Multivariate time series forecasting (MTSF) predicts future values of multiple variables from historical data. While quantum neural networks have been increasingly applied to this task, they typically rely on fixed local measurements, which restrict their expressivity. We propose MTSF-ANO, a simple hybrid model for MTSF that integrates variational quantum circuits with adaptive non-local observables (ANO). On the four ETT datasets, MTSF-ANO ranks first or second in MSE in 17 of 20 settings, improving over the strongest baseline by up to 20\% on ETTh1, and outperforms or matches its fixed local observable counterpart across all settings. Our ablations show how the quantum circuit design and ANO non-locality affect performance. These results suggest that ANO is a promising direction for quantum time series forecasting.
\end{abstract}
\section{Introduction}
Multivariate time series forecasting (MTSF), which predicts multiple variables from historical data, is critical for strategic planning in domains such as energy management, weather modeling, and finance. Notably, the high-dimensional correlations in the data require models to learn complex distributions.

Quantum machine learning (QML) enhances classical machine learning by leveraging the representational expressivity stemming from quantum phenomena such as superposition, entanglement, and quantum interference~\cite{Cerezo2021, McClean_2016}. Among QML frameworks, quantum neural networks (QNNs) are increasingly applied to complex machine learning tasks, such as reinforcement learning~\cite{NEURIPS2021_eec96a7f, 10889145, lee2026quantumhierarchicalreinforcementlearning,Chen:2020opi, Lockwood_Si_2020}, classification~\cite{farhi2018classificationquantumneuralnetworks, PerezSalinas2020datareuploading}, data compression~\cite{Romero_2017}, anomaly detection~\cite{11464824}, and time-series prediction~\cite{qaio25, 9747369, 10650743, 11461293}.

However, QNNs are typically constructed from variational quantum circuits (VQCs) and rely on local measurements, such as Pauli observables. This reliance on local measurements restricts the network's expressivity and its ability to learn complex data distributions. To overcome this bottleneck, recent research suggests jointly optimizing circuit parameters alongside trainable observables~\cite{11250270, 11011001}. Specifically, the adaptive non-local observables (ANO) framework~\cite{11249836} employs trainable multi-qubit Hermitian observables, showing strong potential across super-resolution~\cite{11463496}, reinforcement learning~\cite{11250031}, and classification tasks~\cite{11249836}.

In this work, we propose MTSF-ANO, a simple hybrid model for MTSF that integrates ANO into a data re-uploading VQC~\cite{PerezSalinas2020datareuploading, PhysRevA.103.032430}.\footnote{Code at \url{https://github.com/Yu-TingLee/MTSF-ANO}.} We benchmark MTSF-ANO against strong classical and quantum baselines, and introduce a channel-independent variant that excels at longer lookback windows. We conduct ablations on circuit design and ANO non-locality. Our findings establish ANO-based hybrid models as novel and effective solutions for time-series. Our contributions are:
\begin{itemize}
    \item We introduce MTSF-ANO, a simple hybrid model for MTSF that utilizes trainable non-local observables to enhance forecasting performance.
    \item On the four ETT datasets, MTSF-ANO ranks first or second in 17 out of 20 settings, outperforming or matching the fixed local observable counterpart in all of them.
    \item Ablations show that non-local measurement is the key driver of gains, with entanglement, shallow depth, and moderate non-locality beneficial.
\end{itemize}

\section{Methodology}
\subsection{Problem Formulation}
Consider a multivariate time series dataset with $C$ variates (channels). Let $L$ denote the size of the lookback window and $H$ the forecasting horizon. Given historical data $\mathbf{X}_t \in \mathbb{R}^{C\times L}$, the goal of MTSF is to predict future values $\mathbf{\widehat{Y}}_t \in \mathbb{R}^{C\times H}$. The corresponding ground truth is denoted $\mathbf{Y}_t \in \mathbb{R}^{C\times H}$.

\subsection{Variational Quantum Circuits}
Variational quantum circuits (VQCs), or parameterized quantum circuits (PQCs), are trainable quantum models that process classical data in three stages. First, a data encoding unitary circuit $U(x)$ maps a classical input $x$ into an $n$-qubit system, yielding the encoded states $U(x)|0\rangle^{\otimes n}$, where $|0\rangle^{\otimes n}$ is the ground state. Next, a parameterized unitary circuit $V(\theta)$ evolves the encoded states into $V(\theta)U(x)|0\rangle^{\otimes n}$. This variational circuit $V(\theta)$ typically consists of alternating layers of trainable single-qubit rotations and multi-qubit entangling gates. Finally, a measurement layer is applied to extract classical information by evaluating the expectation values of a fixed Hermitian observable $H$. The computation of a VQC can be summarized as a quantum function $f_{\text{VQC}}(x;\theta)$:
\begin{equation}
f_{\text{VQC}}(x;\theta) = \langle 0|^{\otimes n} U^\dagger(x) V^\dagger(\theta) H V(\theta) U(x) |0\rangle^{\otimes n}.
\end{equation}

\subsection{Adaptive Non-Local Observables}
Adaptive non-local observables (ANO)~\cite{11249836} replace the fixed observable of a generic VQC with a trainable Hermitian $H(\phi)$ parameterized by $\phi$. A $k$-local observable takes the form:
\begin{equation}\label{tag:k-local}
H(\phi) =
\begin{pmatrix}
c_{11} & a_{12} + i b_{12} & a_{13} + i b_{13} & \cdots & a_{1K} + i b_{1K} \\
* & c_{22} & a_{23} + i b_{23} & \cdots & a_{2K} + i b_{2K} \\
* & * & c_{33} & \cdots & a_{3K} + i b_{3K} \\
\vdots & \vdots & \vdots & \ddots & \vdots \\
* & * & * & \cdots & c_{KK}
\end{pmatrix}
\end{equation}
where $k \leq n$, $K = 2^k$, and $\phi = (a_{ij}, b_{ij}, c_{ii})_{i,j=1}^K$ is a set of $K^2$ real parameters.

Making the observable trainable strictly enlarges the function class: a standard VQC with fixed local observables is provably a special case of ANO~\cite{11249836}. Moreover, a $k$-local observable $H(\phi)$ acts jointly on $k$ qubits, coupling features across distant qubits and promoting an information mixture that single-qubit Pauli measurements cannot express. This expressivity makes ANO well-suited to the complex, high-dimensional correlations in MTSF.


\subsection{MTSF-ANO}
\label{sec:mtsf-ano}
MTSF-ANO has three parts: instance normalization, a data
re-uploading VQC (DRVQC) with ANO, and a prediction head.
\begin{figure}[th]
    \centering
\resizebox{\columnwidth}{!}{
\begin{quantikz}[column sep=0.4cm]
    \lstick{$\ket{0}_0$} & \gate{H} & \gate{R_z(\theta_{0,1})} \gategroup[4,steps=6,style={dashed,rounded corners,inner sep=1pt,fill=blue!20,fill opacity=0.3}]{Variational $V(\theta)$} & \gate{R_y(\theta_{0,2})} & \ctrl{1} & \qw & \qw & \targ{} & \gate{R_y(w_{0,1}h_{0})} \gategroup[4,steps=2,style={dashed,rounded corners,inner sep=1pt,fill=green!20,fill opacity=0.3}]{Encoding $U(\textbf{h}, \textbf{w})$} & \gate{R_z(w_{0,2}h_{0})} & \meter{} \gategroup[4,steps=1,style={dashed,rounded corners,inner sep=1pt,fill=red!20,fill opacity=0.3}]{$H(\phi)$} \\
    \lstick{$\ket{0}_1$} & \gate{H} & \gate{R_z(\theta_{1,1})} & \gate{R_y(\theta_{1,2})} & \targ{} & \ctrl{1} & \qw & \qw & \gate{R_y(w_{1,1}h_{1})} & \gate{R_z(w_{1,2}h_{1})} & \meter{} \\
    \lstick{$\ket{0}_2$} & \gate{H} & \gate{R_z(\theta_{2,1})} & \gate{R_y(\theta_{2,2})} & \qw & \targ{} & \ctrl{1} & \qw & \gate{R_y(w_{2,1}h_{2})} & \gate{R_z(w_{2,2}h_{2})} & \meter{} \\
    \lstick{$\ket{0}_3$} & \gate{H} & \gate{R_z(\theta_{3,1})} & \gate{R_y(\theta_{3,2})} & \qw & \qw & \targ{} & \ctrl{-3} & \gate{R_y(w_{3,1}h_{3})} & \gate{R_z(w_{3,2}h_{3})} & \meter{}
\end{quantikz}
}
\caption{\textbf{VQC architecture of MTSF-ANO.} Each layer consists a variational circuit of parameterized $R_z$, $R_y$ rotations, circular CNOT gates, and a data re-uploading encoding $U(\mathbf{h}, \mathbf{w})$ with trainable input scaling $\mathbf{w}$. For measurement, trainable $k$-local observables $H(\phi)$ are employed.}
\label{fig:vqc_architecture}
\end{figure}
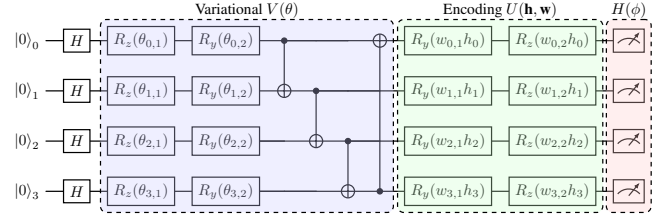
\subsubsection{Instance Normalization}
We employ instance normalization~\cite{kim2022reversible}, which is commonly used in prior forecasting work, to address the potential distribution shift. Given $\mathbf{X}_t \in \mathbb{R}^{C\times L}$, we normalize each channel:
\begin{equation}
  \mathbf{X}^\prime_t = \left(\mathbf{X}_t - \boldsymbol{\mu}\right)\odot{(\boldsymbol{\sigma}^2 + \boldsymbol{\epsilon})^{-1/2}}.
\end{equation}
Here, $\boldsymbol{\mu}, \boldsymbol{\sigma}^2 \in \mathbb{R}^{C}$ are the per-channel mean and variance respectively, $\boldsymbol{\epsilon}$ ensures numerical stability, and $\odot$ denotes element-wise multiplication.

\subsubsection{Data Re-Uploading VQC with ANO}
The normalized input is flattened and projected by a linear layer to a latent representation $\mathbf{h}_t \in \mathbb{R}^n$, where $n$ is the number of qubits. Starting from a layer of Hadamard gates, a DRVQC then transforms $\mathbf{h}_t$. Each VQC layer consists of three parts: parameterized $R_z$ and $R_y$ rotation gates, a circular topology of CNOT gates for entanglement, along with $R_y$ and $R_z$ encoding gates that re-upload the latent input scaled by trainable parameters $\mathbf{w}$ (Fig.~\ref{fig:vqc_architecture}). For measurement, we use combinatorial measurement with $k$-local observables. Specifically, we measure all $\binom{n}{k}$ combinations of $k$ qubits out of the $n$ available, producing an output value per combination and thereby accounting for multi-qubit correlations.
\subsubsection{Prediction Head}
To transform the quantum representations into forecasts, we apply a linear layer followed by de-normalization:
\begin{equation}
\mathbf{\widehat{Y}}_t = \left( W_{\text{out}}\mathbf{f}_{\text{VQC}}(\mathbf{h}_t; \theta, \phi) +\mathbf{b}_{\text{out}}\right) \odot \sqrt{\boldsymbol{\sigma}^2 + \boldsymbol{\epsilon}} + {\boldsymbol\mu},
\end{equation}
where $\mathbf{f}_{\text{VQC}}(\mathbf{h}_t; \theta, \phi) \in \mathbb{R}^{\binom{n}{k}}$ is the vector of expectation values from the quantum circuit.

\section{Experimental Settings}

\begin{table}[th]
\centering
\caption{\textbf{Statistics of the datasets.}}
\label{tab:dataset_stats}
\resizebox{\columnwidth}{!}{
\begin{tabular}{lccc}
\toprule
Datasets & ETTh1 \& ETTh2 & ETTm1 \& ETTm2 \\
\midrule
Variates    & 7   & 7  \\
Timesteps   & 17,420 & 69,680 \\
Sample rate & 1 hour & 5 min \\
\bottomrule
\end{tabular}
}
\end{table}

\begin{table*}[th]
\centering
\caption{\textbf{Multivariate forecasting results.} Results are averaged over 10 runs with different random seeds. Lookback window size $L = 16$ and prediction horizon $H \in \{1, 5, 16, 32, 48\}$. Lower MSE and MAE indicate better performance. The best result is highlighted in \textbf{bold} and the second best is highlighted with \underline{underline}. IMP. is the improvement between MTSF-ANO and the best baseline, where a larger value indicates a better improvement.}
\label{tab:results}
\resizebox{1.0\textwidth}{!}{
\begin{tabular}{c|c||c||cc||cc|cc|cc|cc|cc|cc|cc|cc}
\hline
\multicolumn{2}{c||}{Methods}& \multicolumn{1}{c||}{IMP.}& \multicolumn{2}{c||}{MTSF-ANO} & \multicolumn{2}{c|}{MTSF-PZ} & \multicolumn{2}{c|}{QuLTSF} & \multicolumn{2}{c|}{QLSTM} & \multicolumn{2}{c|}{QFWP} & \multicolumn{2}{c|}{LSTM} & \multicolumn{2}{c|}{DLinear} & \multicolumn{2}{c|}{NLinear} & \multicolumn{2}{c}{Linear} \\
\hline
\multicolumn{2}{c||}{Metric} &MSE & MSE & MAE & MSE & MAE & MSE & MAE & MSE & MAE & MSE & MAE & MSE & MAE & MSE & MAE & MSE & MAE & MSE & MAE\\
\hline
\multirow{5}{*}{ETTh1} 
 & 1 & 15.4\% & \textbf{0.137} & \textbf{0.243} & 0.167 & 0.272 & \underline{0.162} & \underline{0.254} & 0.395 & 0.410 & 0.397 & 0.406 & 0.238 & 0.332 & 0.164 & 0.258 & \underline{0.162} & 0.256 & 0.199 & 0.284 \\
 & 5 & 20.1\% & \textbf{0.338} & \textbf{0.371} & \underline{0.423} & 0.418 & 0.431 & \underline{0.409} & 0.686 & 0.520 & 0.700 & 0.522 & 0.478 & 0.449 & 0.478 & 0.429 & 0.489 & 0.432 & 0.510 & 0.443 \\
 & 16 & 11.2\% & \textbf{0.388} & \textbf{0.405} & 0.453 & 0.439 & \underline{0.437} & \underline{0.422} & 0.607 & 0.493 & 0.662 & 0.514 & 0.469 & 0.447 & 0.459 & 0.428 & 0.487 & 0.441 & 0.474 & 0.435 \\
 & 32 & 11.4\% & \textbf{0.426} & \textbf{0.427} & 0.497 & 0.460 & \underline{0.481} & \underline{0.447} & 0.622 & 0.501 & 0.670 & 0.523 & 0.499 & 0.463 & 0.494 & \underline{0.447} & 0.518 & 0.458 & 0.507 & 0.453 \\
 & 48 & 9.9\% & \textbf{0.427} & \textbf{0.425} & 0.485 & 0.453 & \underline{0.474} & 0.441 & 0.604 & 0.498 & 0.645 & 0.514 & 0.493 & 0.459 & 0.481 & \underline{0.439} & 0.502 & 0.450 & 0.490 & 0.443 \\
\hline
\multirow{5}{*}{ETTh2} 
 & 1 & -6.8\% & 0.079 & 0.173 & 0.090 & 0.191 & \textbf{0.074} & \underline{0.167} & 0.112 & 0.222 & 0.117 & 0.225 & 0.097 & 0.200 & \textbf{0.074} & \textbf{0.166} & \textbf{0.074} & \textbf{0.166} & \underline{0.076} & 0.170 \\
 & 5 & 0.8\% & \textbf{0.121} & \textbf{0.218} & 0.131 & 0.233 & \underline{0.122} & \underline{0.222} & 0.150 & 0.255 & 0.155 & 0.261 & 0.136 & 0.238 & 0.126 & 0.225 & 0.126 & 0.226 & 0.127 & 0.228 \\
 & 16 & -1.7\% & 0.180 & \textbf{0.268} & 0.181 & \underline{0.269} & \textbf{0.177} & 0.271 & 0.192 & 0.283 & 0.193 & 0.284 & 0.183 & 0.272 & \textbf{0.177} & \underline{0.269} & 0.179 & 0.271 & \underline{0.178} & 0.270 \\
 & 32 & -1.4\% & 0.225 & \underline{0.295} & 0.225 & \underline{0.295} & \textbf{0.222} & 0.297 & 0.237 & 0.307 & 0.243 & 0.314 & 0.227 & 0.297 & \textbf{0.222} & \textbf{0.294} & \underline{0.223} & 0.296 & \textbf{0.222} & \underline{0.295} \\
 & 48 & -1.9\% & 0.264 & 0.318 & 0.264 & 0.318 & \textbf{0.259} & 0.318 & 0.273 & 0.327 & 0.277 & 0.331 & 0.265 & 0.319 & \textbf{0.259} & \textbf{0.315} & \underline{0.261} & 0.317 & \textbf{0.259} & \underline{0.316} \\
\hline
\multirow{5}{*}{ETTm1} 
 & 1 & 4.0\% & \textbf{0.048} & \textbf{0.135} & \underline{0.050} & 0.137 & 0.051 & \underline{0.136} & 0.073 & 0.176 & 0.105 & 0.201 & 0.054 & 0.149 & 0.052 & 0.138 & 0.052 & 0.137 & 0.052 & 0.138 \\
 & 5 & 6.8\% & \textbf{0.110} & \textbf{0.204} & \underline{0.118} & \underline{0.210} & 0.128 & \underline{0.210} & 0.182 & 0.266 & 0.198 & 0.271 & 0.119 & 0.216 & 0.133 & 0.214 & 0.132 & 0.214 & 0.133 & 0.214 \\
 & 16 & 4.7\% & \textbf{0.328} & \textbf{0.342} & 0.370 & 0.360 & 0.432 & 0.369 & 0.456 & 0.400 & 0.520 & 0.423 & \underline{0.344} & \underline{0.354} & 0.451 & 0.378 & 0.451 & 0.378 & 0.452 & 0.378 \\
 & 32 & 2.0\% & \textbf{0.594} & \textbf{0.462} & 0.669 & 0.492 & 0.788 & 0.517 & 0.771 & 0.531 & 0.859 & 0.552 & \underline{0.606} & \underline{0.476} & 0.835 & 0.534 & 0.836 & 0.534 & 0.836 & 0.534 \\
 & 48 & -1.8\% & \underline{0.720} & \textbf{0.522} & 0.808 & 0.553 & 0.956 & 0.588 & 0.913 & 0.590 & 0.986 & 0.608 & \textbf{0.707} & \underline{0.525} & 1.018 & 0.610 & 1.019 & 0.610 & 1.019 & 0.610 \\
\hline
\multirow{5}{*}{ETTm2} 
 & 1 & -6.2\% & \underline{0.034} & 0.106 & 0.035 & 0.106 & \textbf{0.032} & \textbf{0.095} & 0.044 & 0.127 & 0.051 & 0.136 & 0.037 & 0.113 & \textbf{0.032} & \textbf{0.095} & \textbf{0.032} & \underline{0.096} & \textbf{0.032} & \underline{0.096} \\
 & 5 & -1.7\% & \underline{0.059} & \underline{0.141} & 0.060 & 0.144 & \textbf{0.058} & \textbf{0.138} & 0.071 & 0.164 & 0.073 & 0.166 & 0.062 & 0.148 & 0.060 & 0.142 & 0.060 & 0.142 & 0.060 & 0.142 \\
 & 16 & 2.9\% & \textbf{0.100} & \textbf{0.191} & \underline{0.103} & \underline{0.196} & 0.104 & \underline{0.196} & 0.108 & 0.202 & 0.118 & 0.216 & 0.104 & 0.197 & 0.108 & 0.201 & 0.108 & 0.201 & 0.108 & 0.201 \\
 & 32 & 4.6\% & \textbf{0.146} & \textbf{0.238} & \underline{0.153} & \underline{0.245} & 0.158 & 0.250 & 0.161 & 0.253 & 0.168 & 0.263 & 0.154 & 0.246 & 0.162 & 0.255 & 0.162 & 0.255 & 0.162 & 0.255 \\
 & 48 & 3.8\% & \textbf{0.179} & \textbf{0.267} & \underline{0.186} & \underline{0.274} & 0.195 & 0.284 & 0.196 & 0.283 & 0.205 & 0.295 & \underline{0.186} & \underline{0.274} & 0.198 & 0.288 & 0.198 & 0.288 & 0.198 & 0.288 \\
\hline
\end{tabular}
}
\end{table*}

\subsection{Dataset}
We conduct experiments on four widely-used real-world datasets from the Electricity Transformer Temperature (ETT) benchmark~\cite{Zhou_Zhang_Peng_Zhang_Li_Xiong_Zhang_2021}. Statistics of the datasets are reported in Table~\ref{tab:dataset_stats}. Following standard practice, we split each dataset into training, validation, and test sets with a 6:2:2 ratio.

\subsection{Evaluation Metrics}
Following prior work~\cite{qaio25,Zhou_Zhang_Peng_Zhang_Li_Xiong_Zhang_2021}, we report mean squared error (MSE) and mean absolute error (MAE), calculated as $\text{MSE} = \frac{1}{C \cdot H} \| \mathbf{Y}_t - \mathbf{\widehat{Y}}_t \|_F^2$ and $\text{MAE} = \frac{1}{C \cdot H} \| \mathbf{Y}_t - \mathbf{\widehat{Y}}_t \|_1$. Here, $\|\cdot\|_F$ and $\|\cdot\|_1$ denote the Frobenius and the $L^1$-norm, respectively.

\subsection{Baselines}
\label{sec:baselines}

We benchmark MTSF-ANO against several classical and quantum baselines. Following prior works, we evaluate classical and quantum recurrent networks: LSTM, quantum LSTM (QLSTM)~\cite{9747369}, and quantum fast weight programmers (QFWP)~\cite{10650743}. We also include state-of-the-art linear methods DLinear, NLinear, and Linear~\cite{Zeng_Chen_Zhang_Xu_2023}, which are strong baselines known to outperform many Transformer-based methods in long-term MTSF. We further compare against QuLTSF~\cite{qaio25}, a state-of-the-art quantum method for long-term MTSF. Finally, to isolate the effect of ANO measurement, we include an MTSF-PZ baseline that replaces the $k$-local measurement in MTSF-ANO with fixed Pauli-$Z$ observables.

The LSTM baseline uses 3 layers with 7 hidden units. QLSTM, QFWP, QuLTSF, and the three linear baselines are rerun using their official code. For a fair comparison, all recurrent models use a linear prediction head for direct multi-step forecasting, and all baselines utilize instance normalization.

\subsection{Hyperparameters}
For all quantum methods, the number of qubits equals to the channel count $C$, and the circuit depth is 3 layers. We train for 100 epochs with Adam (learning rate 0.001) using the MSE loss. Early stopping is applied with a patience of 10 epochs.

\section{Experiments}


\subsection{Main Results}
\begin{figure}[th!]
    \centering
    \includegraphics[width=\columnwidth]{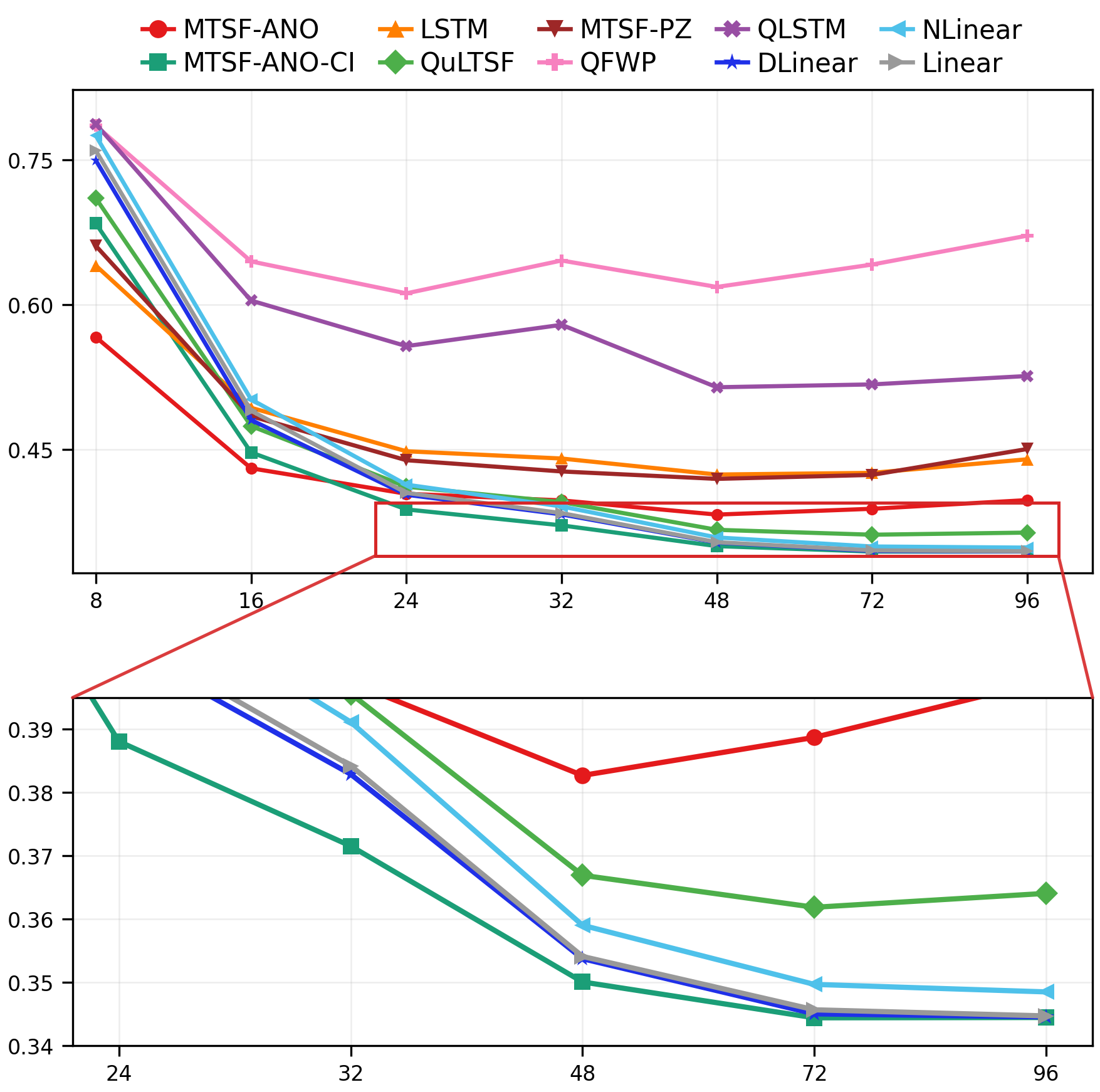}
    \caption{\textbf{Impact of lookback window size (ETTh1).} We report MSE with $H = 48$ and varying lookback from $L = 8$ to $L = 96$. MTSF-ANO and its channel-independent variant, MTSF-ANO-CI, outperform all other models, with MTSF-ANO best at $L \in \{8, 16\}$ and MTSF-ANO-CI best for $L > 16$.}
    \label{fig:ablation_lookback}
\end{figure}

Table~\ref{tab:results} summarizes the forecasting performance at a fixed lookback $L = 16$, reporting the best result over $k \in \{1, \dots, 7\}$ for each setting. Fig.~\ref{fig:ablation_lookback} reports MSE as the lookback window varies at $H = 48$ on ETTh1, likewise reporting the best result over $k \in \{3, 5\}$. MTSF-ANO ranks first or second in terms of MSE in 17 of the 20 settings. Its advantage is largest on ETTh1, where it improves over the strongest baseline by 9–20\%, and it remains competitive on ETTm1 and ETTm2. This advantage narrows as the lookback $L$ grows, with the linear baselines starting to surpass MTSF-ANO. This is likely because projecting a longer input down to $n$ qubits becomes a bottleneck. To address it, we further introduce a channel-independent variant, MTSF-ANO-CI, whose input and output projections use per-channel shared-weight linear layers. As Fig.~\ref{fig:ablation_lookback} shows, MTSF-ANO and MTSF-ANO-CI are the strongest models for $L\in\{8, 16\}$ and $L = 24$ to $L = 96$, respectively. Finally, the comparisons against the MTSF-PZ baseline confirm that ANO effectively enhances MTSF capabilities (Table~\ref{tab:drvqc_imp}).

\begin{table}[th]
\centering
\caption{\textbf{Improvement over MTSF-PZ.} We report the IMP. of MTSF-ANO over MTSF-PZ from Table~\ref{tab:results}. Positive values indicate MTSF-ANO is better.}
\label{tab:drvqc_imp}
\resizebox{\columnwidth}{!}{
\begin{tabular}{lccccc}
\toprule
Settings & $H = 1$ & $H = 5$ & $H = 16$ & $H = 32$ & $H = 48$ \\
\midrule
ETTh1    & 18.0\% & 20.1\% & 14.3\% & 14.3\% & 12.0\% \\
ETTh2    & 12.2\% & 7.6\% & 0.6\% & 0.0\% & 0.0\% \\
ETTm1    & 4.0\% & 6.8\% & 11.4\% & 11.2\% & 10.9\% \\
ETTm2    & 2.9\% & 1.7\% & 2.9\% & 4.6\% & 3.8\% \\
\bottomrule
\end{tabular}
}
\end{table}

\subsection{Impact of ANO Non-Locality}
Table~\ref{tab:ablation_locality} reports the improvement over MTSF-PZ, averaged across the ETT datasets, as the non-locality $k$ varies. Performance increases steadily from $k = 1$ and reaches its peak between $k = 3$ and $k = 5$, where MTSF-ANO demonstrates a 6.4\% to 8.5\% average improvement over MTSF-PZ. Interestingly, the gains collapse sharply at $k = 7$. We attribute this to our combinatorial ANO scheme: at $k=7$, only a single expected value is evaluated and transformed into forecasts. These results also indicate that a moderate non-locality $k$ is the sweet spot for our propose model.

\begin{table}[t!]
\centering
\caption{\textbf{Impact of ANO non-locality.} We report the IMP. over MTSF-PZ, averaged across the four ETT datasets, as the non-locality $k$ varies. \textbf{Bold} indicates the best $k$ for each $H$.} 
\label{tab:ablation_locality}
\resizebox{\columnwidth}{!}{
\begin{tabular}{cccccc}
\toprule
Settings  & $H = 1$ & $H = 5$ & $H = 16$ & $H = 32$ & $H = 48$ \\
\midrule
$k=1$  & 0.3\%           & 0.2\%          & -0.3\%             & -0.1\%         & -0.7\% \\
$k=2$  & 5.4\%           & 4.8\%          & 3.1\%              & 3.7\%          & 3.3\% \\
$k=3$  & 7.4\%           & \textbf{8.5}\% & 5.4\%              & 5.7\%          & 5.2\% \\
$k=4$  & 7.2\%           & 7.5\%          & 6.5\%              & 6.8\%          & \textbf{6.4}\% \\
$k=5$  & \textbf{8.0}\%  & 7.8\%          & \textbf{7.3}\%     & \textbf{7.4}\% & 6.3\% \\
$k=6$  & 7.2\%           & 7.1\%          & 5.4\%              & 5.1\%          & 3.6\% \\
$k=7$  & -97.5\% & -39.7\% & -24.9\% & -19.4\% & -18.0\% \\
\bottomrule
\end{tabular}
}
\end{table}

\begin{table}[th]
\centering
\caption{\textbf{Impact of VQC architectures (ETTh1).} We report MSE with and without trainable input scaling (left $\mid$ right) and CNOT gates, and varying non-locality $k$ and circuit depth. }
\label{tab:ablation_architecture}

\begin{subtable}{\columnwidth}
\centering
\caption{$H = 1$}
\label{tab:ablation_architecture_h1}
\resizebox{\columnwidth}{!}{
\begin{tabular}{ccccc}
\toprule
Architecture & \multicolumn{2}{c}{$k=3$} & \multicolumn{2}{c}{$k=5$} \\
\cmidrule(lr){2-3} \cmidrule(lr){4-5}
Layer & w/ CNOT & w/o CNOT & w/ CNOT & w/o CNOT \\
\midrule
1 & \textbf{0.137} $\mid$ 0.138 & 0.138 $\mid$ 0.139 & 0.139 $\mid$ 0.139 & \textbf{0.138} $\mid$ 0.140 \\
3 & \textbf{0.141} $\mid$ 0.142 & 0.142 $\mid$ 0.142 & \textbf{0.137} $\mid$ 0.138 & 0.143 $\mid$ 0.142 \\
5 & \textbf{0.142} $\mid$ 0.145 & 0.153 $\mid$ 0.153 & \textbf{0.139} $\mid$ 0.140 & 0.172 $\mid$ 0.175 \\
\bottomrule
\end{tabular}
}
\end{subtable}

\vspace{1em}

\begin{subtable}{\columnwidth}
\centering
\caption{$H = 48$}
\label{tab:ablation_architecture_h48}
\resizebox{\columnwidth}{!}{
\begin{tabular}{ccccc}
\toprule
Architecture & \multicolumn{2}{c}{$k=3$} & \multicolumn{2}{c}{$k=5$} \\
\cmidrule(lr){2-3} \cmidrule(lr){4-5}
Layer & w/ CNOT & w/o CNOT & w/ CNOT & w/o CNOT \\
\midrule
1 & \textbf{0.430} $\mid$ 0.431 & 0.432 $\mid$ 0.433 & 0.431 $\mid$ \textbf{0.430} & 0.433 $\mid$ 0.433 \\
3 & 0.437 $\mid$ \textbf{0.436} & 0.440 $\mid$ 0.439 & 0.431 $\mid$ \textbf{0.430} & 0.441 $\mid$ 0.441 \\
5 & \textbf{0.436} $\mid$ 0.440 & 0.453 $\mid$ 0.454 & \textbf{0.431} $\mid$ \textbf{0.431} & 0.462 $\mid$ 0.459 \\
\bottomrule
\end{tabular}
}
\end{subtable}

\end{table}

\subsection{Impact of VQC Architectures}
Table~\ref{tab:ablation_architecture} ablates three architectural choices and reports MSE at $H \in \{1, 48\}$ on ETTh1.
First, entanglement helps: removing the CNOT gates (w/o CNOT) degrades performance in nearly every setting, and the gap widens at higher non-locality and greater depth. Second, the trainable input scaling $\mathbf{w}$ yields marginally better results in most settings, but the difference is negligible. Third, compared to the default 3 layers, decreasing to 1 layer generally decreases MSE, while increasing layers generally increases it. Overall, entanglement and shallow depth stand out as the most beneficial architectural factors.

\section{Conclusion}
This work introduced MTSF-ANO, a hybrid model that integrates adaptive non-local observables into a VQC for multivariate time series forecasting. MTSF-ANO utilizes trainable non-local measurements and yields the best or second-best MSE in 17 of 20 settings across the four ETT datasets, surpassing or matching its fixed local observable counterpart in all settings. A channel-independent variant further extends these gains to longer lookback windows. Our ablations identify non-local measurement as the main driver of these gains, with entanglement and shallow depth also beneficial and non-locality showing a sweet spot. Our results establish ANO as a promising direction for quantum time series forecasting.



\vfill\pagebreak

\ninept
\bibliographystyle{IEEEbib}
\bibliography{reference}

\end{document}